\documentclass[12pt]{article}
\usepackage{graphicx}
\usepackage{amsmath,amssymb,amsfonts}
\usepackage[numbers,sort&compress]{natbib}
\usepackage[colorlinks=true, linkcolor=blue, urlcolor=blue, citecolor=blue]{hyperref}
\usepackage{bm}
\usepackage{tabularx}
\usepackage{xcolor}
\usepackage{booktabs}
\usepackage{caption,subcaption}
\usepackage{authblk}
\usepackage{tabularx}
\usepackage{float}

\title{\textbf{A Study of Non-Singular Bounce in Myrzakulov-type $f(R,T)$ Gravity with Chaplygin Gas}}

\author[1]{Khandro Kalsang\thanks{\href{mailto:khankalmaths2023@outlook.com}{khankalmaths2023@outlook.com}}}
\author[2,3,4]{Abdel Nasser Tawfik\thanks{\href{mailto:a.tawfik@fue.edu.eg}{a.tawfik@fue.edu.eg}}}
\author[1]{Surajit Chattopadhyay\thanks{* Corresponding author: 
\href{mailto:surajitchatto@outlook.com}{surajitchatto@outlook.com}, 
\href{mailto:schattopadhyay1@kol.amity.edu}{schattopadhyay1@kol.amity.edu}}}

\affil[1]{\small Department of Mathematics, Amity University Kolkata, Kolkata 700135, India}
\affil[2]{\small Ahram Canadian University, Faculty of Engineering, Basic Science Department, 12573 Giza, Egypt}
\affil[3]{\textcolor{black}{Physics Department, Faculty of Science, Islamic University of Madinah, 13518 Madinah, KSA}}
\affil[4]{\small The Egyptian Center for Theoretical Physics (ECTP), Giza, Egypt}
\date{\today}

\begin{document}

\maketitle

\begin{abstract}
This study investigates the non-singular bounce within the framework of Myrzakulov-type $f(R,T) = R + \alpha T + \beta T^2$ gravity by adopting a Chaplygin gas equation of state. We employ two methodologies: a reconstruction scheme via a symmetric scale factor ansatz (Model I) and an autonomous dynamical system analysis (Model II). Our results indicate that the quadratic trace parameter $\beta$ acts as a primary physical driver; specifically, for $\beta < 0$, the matter-geometry coupling generates sufficient geometric repulsion to effectively violate the Null Energy Condition (NEC) at high densities without the requirement of exotic matter fields. A numerical scan of the $(\beta, \rho_0)$ parameter space indicates a critical density threshold required to initiate the bounce, below which the Universe follows a singular General Relativity trajectory. The models are shown to be physically viable, with the effective equation of state asymptotically approaching a de Sitter attractor ($w_{\text{eff}} \to -1$) and the squared speed of sound remaining within the stability and causality bounds ($0 \le c_s^2 \le 1$). This study shows that the $f(R,T)$ framework provides a stable, classically geometric alternative to the Big Bang singularity, consistent with both early-universe requirements and late-time accelerated expansion.
\end{abstract}

\tableofcontents

\section{Introduction}
General Relativity (GR) and the remarkably effective $\Lambda$CDM model \citep{del2017small, odintsov2025modified, blanchard2022lambda} have been dominating the evolution of the modern Universe. They collectively explain a Universe which started from an extremely dense and hot condition and has been expanding for billions of years. Nevertheless, the standard Big Bang model, with all its power to predict, is still essentially unfinished \citep{lal2010big}. Penrose and Hawking's singularity theorems \citep{hawking1970singularities} imply that a Universe expanding according to GR and comprising standard matter has to come from a singularity, i.e. an infinite density and spacetime curvature point where the existing physics laws break down. This "initial singularity problem" indicates that GR might only be an effective theory rather than an ultimate one, especially in the highly energetic regimes of the early Universe. Hence, theoretical physicists have been looking for alternatives \citep{chamseddine2017resolving} that would eliminate the singularity entirely.
The most interesting of the alternatives, perhaps, is the non-singular bounce cosmology \citep{novello2008bouncing}. In such a scenario, the Universe does not start at a point of singularity; rather, it experiences a smooth change from a previous contraction phase to the present expansion phase. When a cosmological bounce takes place, the Hubble parameter $H$ becomes zero and its time derivative $ \dot{H}$ stays positive: $H=0, \quad \dot{H}>0$. The Raychaudhuri equation of standard GR \citep{kar2007raychaudhuri} prohibits such a change of phase if the matter content is such that the Null Energy Condition (NEC): $\rho + p \ge 0$ holds. Therefore, in order to for bounce to occur, the introduction of exotic matter fields or the alteration of the gravity's geometric sector is necessary.
Modified theories of gravity, by offering ghost-free geometric corrections, permit the possibility of effectively violating the NEC without running into the instabilities of the ghost fields \citep{libanov2016generalized}.
 In this context, we would like to note the work done by Mironov \textit{et. al.} \citep{mironov2024non}. In their work, the authors have reviewed the development and current state of research on the early universe models without an initial singularity, namely the cosmological bounce and genesis scenarios, within the background of a broad class of scalar-tensor theories, namely the Horndeski theories and their generalizations. Various works on bounce within various contexts can be found in the literature \citep{bajardi2020bouncing, miranda2022bouncing, moriconi2016big, odintsov2015bouncing, odintsov2014matter, odintsov2022bounce,saha2023realization, chattopadhyay2023cosmological}. Among the different modified gravity theories like $f(R)$ \citep{sotiriou2010f, bertolami2008f, capozziello2008massive}, $f(T)$ \citep{capozziello2011cosmography, paliathanasis2016cosmological, cai2018f, myrzakulov2011accelerating} gravity theories etc., $f(R,T)$ gravity has emerged as a particularly vesatile thory.  The $f(R,T)$ gravity was first introduced by Harko \textit{et. al.} \citep{harko2011f} where $R$ serves as the gravitational Lagrangian and $T$ is the trace of the stress-energy tensor. In their study, the authors derived the field equations using the metric formalism and showed that the equations of motion for test particles are derived from the covariant divergence of the stress-energy tensor. In addition to this, the authors detail the analysis of scalar-field models within this framework. This feature is particularly relevant for bounce scenarios, as the $T$-dependent terms can effectively provide the repulsive pressure required to bypass the initial singularity without the need for quintom or phantom matter fields \citep{shabani2014cosmological, sharif2014dynamical}. For a more detailed study on the geometric root of $f(R,T)$ gravity within FRW cosmology, one can look into the work of R. Myrzakulov in \citep{myrzakulov2012frw}. Other works on $f(R,T)$ gravity include \citep{das2017gravastars, barrientos2014comment, alvarenga2012testing, sharif2013analysis, fisher2019reexamining, alvarenga2013dynamics, deb2018strange}. Building upon this foundation, recent developments have focused on specific functional forms that can accommodate both early-time inflation and late-time acceleration. One such promising direction is the Myrzakulov-type $f(R,T)$ gravity \citep{myrzakulov2012frw}, which generalize the coupling between geometry and matter through various formalisms, including metric-affine \citep{myrzakulov2012metric, myrzakulov2021metrica}, vielbein, and Weyl-Cartan space-times \citep{momeni2025metric, momeni2025myrzakulov}. These theories have gained significant traction due to their ability to address fundamental cosmological challenges, such as the $H_0$ tension \citep{aljohani2025toward} and baryon asymmetry \citep{saleem2023viable}, while providing a robust platform for gravitational wave \citep{MOMENI2025116903, momeni2025metric} and black hole studies \citep{momeni2025wald, Chen_2018}.The Myrzakulov gravity (MG) landscape is extensive, encompassing versions that incorporate Gauss-Bonnet terms, boundary scalars, and non-metricity \citep{iosifidis2022metric,myrzakul2021metricm,momeni2025cosmologicalc,momeni2025einstein}. Extensive observational constraints have been placed on these models using recent datasets \citep{anagnostopoulos2021observational, kazempour2025cosmological, momeni2025dark}, confirming their viability as alternatives to the standard model. Furthermore, dynamical system analyses \citep{papagiannopoulos2022dynamical} and exact cosmological solutions \citep{saridakis2020cosmological, maurya2024exact, maurya2024transit} have highlighted the rich phenomenology of MG models, particularly in describing the transition from early-time inflation to late-time dark energy dominance \citep{maurya2024flrw, maurya2024metric, momeni2025inflation, lymperis2025correspondence, shaily2026late}. Of particular interest to our study is the capacity of Myrzakulov-type $f(R,T)$ gravity to resolve early-universe singularities. While non-singular bounce solutions have been explored in various modified contexts, such as ekpyrotic scenarios and wormhole physics \citep{bubuianu2024dark, rauf2025exploring, maurya2025quintom}, recent investigations have specifically pointed toward the efficiency of the $f(R,T)$ branch of MG in producing stable, non-singular transitions \citep{lymperis2025nonsingularbouncesolutionsmyrzakulov}. In this work, we investigate a specific $f(R,T)$ model defined by the functional form: $f(R,T) = R + \alpha T + \beta T^2$ where the gravitational action depends on both the Ricci scalar $R$ and the trace of the energy-momentum tensor $T$. The quadratic $T^2$ term is a high-density regulator, which we analyze, through both numerical reconstruction and dynamical system analysis, to determine if it brings about a smooth cosmological bounce.
To model the matter sector, we use the Chaplygin gas equation of state, which is known for combining dark matter and dark energy behaviour. Also, it keeps the standard condition $\rho + p > 0$. Because of this, the violation of the NEC \citep{tawfik2017flrw}  needed for the bounce comes only from the geometric correction in modified gravity. Compared to scalar-field or phantom-based models, the Chaplygin gas avoids the introduction of additional degrees of freedom or ghost instabilities while still capturing the essential features of cosmic evolution \cite{bento2002chaplygin,bento2003cmb,bento2003grg,kamenshchik2000chaplygin}. 

\subsection*{Motivation for the adopted mathematical framework}
The standard cosmological model based on GR predicts an initial Big Bang singularity where physical quantities diverge. This indicates that GR may not be sufficient to describe the early high-energy phase of the Universe. Therefore, modified theories of gravity are often considered to address this problem. Among them, $f(R,T)$ gravity is interesting because the gravitational action depends not only on the Ricci scalar $R$ but also on the trace $T$ of the energy--momentum tensor. This introduces a coupling between matter and geometry which can significantly modify the cosmological dynamics at high densities. In the present work we consider the simple form $f(R,T)=R+\alpha T+\beta T^2$. The quadratic term becomes important in the early Universe and can produce an effective repulsive contribution in the modified field equations. This feature allows the possibility of a non-singular cosmological bounce without introducing exotic matter fields.
Furthermore, it should be mentioned that the Chaplygin gas is adopted in this study due to its unique property of interpolating between a pressureless fluid at early times and a negative-pressure dark energy component at late times, thereby offering a unified framework for cosmic evolution.

In this way, we structure our paper as follows: In Section~ \ref{2}, we define the theoretical basis of the expression of our gravity, $f(R,T)=R+\alpha~T+\beta~T^2$, and formulate the Friedmann equations modified to an FRW Universe. We also give the analytical conditions needed to have a non-singular bounce, including the quadratic trace parameter $\beta$. The main body of our research is provided in Section~\ref{3} in which we consider two methodologies: a kinematic reconstruction by employing a symmetric scale factor representation (Model I) and an autonomous dynamical system representation (Model II). In this part, we analyze the physical viability of the models by considering classical conditions of their energy, stability test in the form of squared speed of sound, $c_s^2$ and using phase space characterization of the world. In Section~\ref{4}, we give a comparative analysis of these two models and the interpretation of our findings in the light of the known bouncing cosmologies. Section~\ref{5} eventually wraps up the paper with a conclusion on our findings and what they imply for the early universe physics.

\section{Theoretical Framework of \texorpdfstring{$f(R,T)$}{f(R,T)} Gravity}\label{2}
Having discussed the motivations in the previous section, we now establish the mathematical foundation for our study. In this section, we outline the metric formalism of $f(R,T)$ gravity. Here the standard Einstein-Hilbert action is generalized to include a functional dependence on the trace of the energy-momentum tensor. This coupling introduces a non-trivial interaction between the geometric curvature and the matter content, which results in a modified set of field equations that deviate from GR in high-curvature regimes.
The gravitational action in $f(R,T)$ gravity is defined as \citep{harko2011f}
\begin{equation}
    S = \int d^4x \sqrt{-g} \left[ \frac{1}{16\pi G} f(R,T) + \mathcal{L}_m \right],
\end{equation}
where $g$ is the determinant of the metric tensor and $\mathcal{L}_m$ is the matter Lagrangian. Varying the action with respect to the metric $g_{\mu\nu}$, we get
\begin{equation}
\begin{array}{cc}
     &f_R(R,T) R_{\mu\nu} - \frac{1}{2} f(R,T) g_{\mu\nu} + (g_{\mu\nu} \Box - \nabla_\mu \nabla_\nu) f_R(R,T)  \\
     &= 8\pi G T_{\mu\nu} - f_T(R,T) (T_{\mu\nu} + \Theta_{\mu\nu}).
\end{array}
\end{equation}
Here, $f_R$ and $f_T$ denote the partial derivatives of $f(R,T)$ with respect to $R$ and $T$, respectively. The tensor $\Theta_{\mu\nu}$ is defined by the variation of the energy-momentum tensor and takes the form $\Theta_{\mu\nu} = -2T_{\mu\nu} + p g_{\mu\nu}$ for a perfect fluid.

In our work, we assume a spatially flat Friedmann-Robertson-Walker (FRW) metric 
\begin{equation}
   ds^2 = -dt^2 + a^2(t) d\mathbf{x}^2, 
\end{equation} 
the modified Friedmann equations for our specific model $f(R,T) = R + \alpha T + \beta T^2$ \citep{Moraes2016} are derived as
\begin{equation}
    3H^2 = \rho + (\alpha + 2\beta T)(\rho + p) + \frac{1}{2}(\alpha T + \beta T^2) = \rho_{\text{\text{eff}}},
\end{equation}
\begin{equation}
    2\dot{H} + 3H^2 = -p + \frac{1}{2}(\alpha T + \beta T^2) = -p_{\text{\text{eff}}}.
\end{equation}
By combining these, we obtain the Raychaudhuri equation, which governs the acceleration and potential bounce of the Universe
\begin{equation}
    \dot{H} = -\frac{1}{2} (\rho_{\text{\text{eff}}} + p_{\text{\text{eff}}}) = -\frac{1}{2}(\rho + p)(1 + \alpha + 2\beta T).
\end{equation}

We consider the matter content as a Chaplygin gas with the equation of state \citep{gorini2003can}
\begin{equation}
    p = -\frac{A}{\rho^\gamma},
\end{equation}
where A is a constant with
dimensions $[M^{4(1+\alpha)}]$. Thus, we observe that the trace of the energy-momentum tensor $T = \rho - 3p$ then becomes a function of the density, $T = \rho + 3A\rho^{-\gamma}$. To analyze the global dynamics, we define a 2D autonomous system in the variables $(H, \rho)$
\begin{align}
    \dot{H} &= -\frac{1}{2} \left( \rho - \frac{A}{\rho^\gamma} \right) \left[ 1 + \alpha + 2\beta \left( \rho + \frac{3A}{\rho^\gamma} \right) \right] \label{dyn1} ,\\
    \dot{\rho} &= -3H \left( \rho - \frac{A}{\rho^\gamma} \right) \label{dyn2}.
\end{align}
The fixed points of this system correspond to $\dot{H} = 0$ and $\dot{\rho} = 0$. A critical point occurs at $\rho^* = A^{1/(1+\gamma)}$, which represents the de Sitter attractor where $p = -\rho$. 

\subsection{Analytical Bounce Condition}\label{s2}
In order to understand the physical mechanism of the singularity resolution, we carry out an analytical derivation of the conditions necessary to have non-singular bouncing. In a Friedmann-Lemaitre-Robertson-Walker (FLRW) cosmology that has a gravitational field and is governed by GR, the Raychaudhuri equation is expressed as $\dot{H} = -4\pi G(\rho + p)$. To have a bounce at $t=0$, we must have the conditions, $H=0$ and $H>0$. When ordinary matter follows the NEC (in GR), a bounce is strictly forbidden as $\dot{H}$ would always be non-positive. For $f(R,T)$ gravity, the effective Raychaudhuri equation is modified by the inclusion of the trace of the energy-momentum tensor. Beginning with the field equations, it can be seen that the Hubble parameter evolves dynamically and obeys
\begin{equation}
\dot{H} = -\frac{1}{2}(\rho_{\text{eff}} + p_{\text{eff}}).
\end{equation}
Substituting the effective energy density $\rho_{\text{eff}}$ and effective pressure $p_{\text{eff}}$
\begin{align}\rho_{\text{eff}} &= \rho + (\alpha + 2\beta T)(\rho + p) + \frac{1}{2}(\alpha T + \beta T^2) ,& p_{\text{eff}} = p - \frac{1}{2}(\alpha T + \beta T^2).\end{align}
We compute the sum $(\rho_{\text{eff}} + p_{\text{eff}})$ to find the slope of the Hubble parameter
\begin{equation}
\rho_{\text{eff}} + p_{\text{eff}} = (\rho + p) + (\alpha + 2\beta T)(\rho + p).
\end{equation}
Factoring out the term $(\rho + p)$, the modified Raychaudhuri equation becomes
\begin{equation}
\dot{H} = -\frac{1}{2}(\rho + p) \left[ 1 + \alpha + 2\beta T \right].
\end{equation}
To satisfy the bounce condition $\dot{H} > 0$ at the point where $H=0$, the right-hand side of the equation must be strictly positive. This gives us
\begin{equation}
(\rho + p)(1 + \alpha + 2\beta T) < 0.
\end{equation}
For a Universe filled with Chaplygin gas, the matter sector generally satisfies the condition $\rho + p > 0$. Consequently, the bounce can only be triggered if
\begin{equation}
1 + \alpha + 2\beta T < 0.
\end{equation}
Substituting the trace for Chaplygin gas $T = \rho + 3A\rho^{-\gamma}$, we get
\begin{equation}
1 + \alpha + 2\beta \left( \rho + \frac{3A}{\rho^\gamma} \right) < 0.
\end{equation}
This derivation shows that the quadratic coupling parameter $\beta$, is the main physical driver. Particularly, in the case of a negative value of the $\beta$, the other term, which is $2\beta T$ is dominant at higher densities, which offers a repulsive gravitational source that essentially violates the NEC and allows the non-singular transition.
In a Universe with Chaplygin gas, the matter component meets the requirement of $\rho + p > 0$. The benefit of this is that the generic matter condition is preserved and the transgression of the NEC required to make the bounce is only due to the geometric component of the $f(R,T)$ gravity and not due to an exotic matter.

\section{Cosmological Dynamics and Model Reconstruction}\label{3}

As we have already determined the theoretical requirements of a bounce in the last section, we now proceed to take two different options to investigate the presence of a non-singular bounce in $f(R,T) = R + \alpha T + \beta T^2$ gravity. To start with, we apply a reconstruction method with a certain scale factor ansatz. Second, we develop the dynamics of the bounce as an autonomous dynamical system to check the stability and robustness.

\subsection{Model I: Reconstruction via Scale Factor Ansatz}

For this model, we consider a scale factor that describes a symmetric transition from a contracting phase to an expanding one. We assume the following ansatz
\begin{equation}
    a(t) = a_0 (1 + t^2)^{h/2},
    \label{scale}
\end{equation}
where $a_0$ is the scale factor at the bounce point ($t=0$) and $h$ is a positive constant governing the rate of expansion. The resulting Hubble parameter $H(t)$ and its time derivative $\dot{H}(t)$ are
\begin{equation}
    H(t) = \frac{ht}{1+t^2}, \quad \dot{H}(t) = \frac{h(1-t^2)}{(1+t^2)^2}~~.
\end{equation}
At the bounce point $t=0$, we find $H(0)=0$ and $\dot{H}(0)=h > 0$, satisfying the necessary conditions for a non-singular bounce.

Before proceeding further, we would like to mention that the choice of scale factor as given in Eq.~(\ref{scale}) is motivated by both physical and mathematical considerations relevant to non-singular bouncing cosmologies. Firstly, this form ensures the realization of a non-singular bounce at $t=0$. In particular, the scale factor remains finite and non-vanishing, $a(0)=a_0 \neq 0$, thereby avoiding the Big Bang singularity. Moreover, the Hubble parameter satisfies the necessary bounce conditions $H(0)=0$ and $\dot{H}(0)=h>0$. Thus, the ansatz provides a smooth transition from a contracting phase ($t<0$) to an expanding phase ($t>0$). 

\begin{figure}[H]
    \centering
    \includegraphics[width=\linewidth]{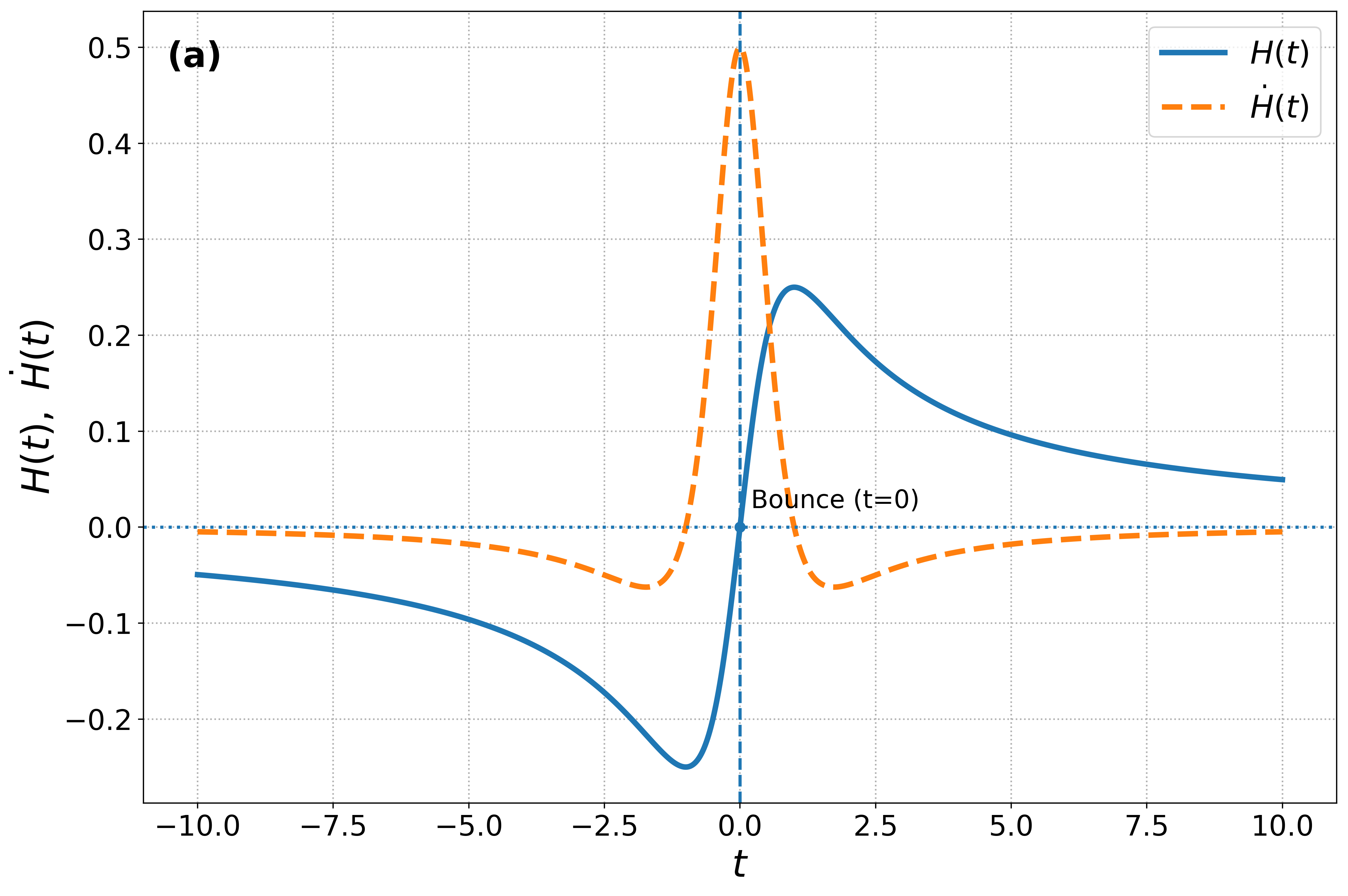}
    \caption{Evolution of the Hubble parameter $H(t)$ and its time derivative $\dot{H}(t)$ for the reconstructed scale factor ansatz $a(t) = a_0(1+t^2)^{h/2}$ with $h = 0.5$.}
    \label{f1}
\end{figure}

\textcolor{black}{Fig.~\ref{f1} illustrates the time evolution of the Hubble parameter $H(t)$ and its derivative $\dot{H}(t)$ around the bouncing point. It can be clearly seen that $H(t)$ changes its sign from negative to positive at $t=0$, indicating the transition of the Universe from a contracting phase to an expanding phase. At the same time, we observe that $\dot{H}(t)$ remains positive near the bouncing point. This satisfies the necessary condition for the occurrence of a non--singular cosmological bounce. Furthermore, from this behaviour it may be interpreted that the adopted ansatz for scale factor produces a smooth bounce without any singular behaviour.}

We consider a Universe filled with a Chaplygin gas. By solving the energy conservation equation $\dot{\rho} + 3H(\rho+p) = 0$ with $p = -A/\rho^\gamma$, the energy density as a function of the scale factor is obtained as
\begin{equation}
    \rho(a) = \left[ A + \frac{B}{a^{3(1+\gamma)}} \right]^{\frac{1}{1+\gamma}}.
\end{equation}
Here, $B$ is the integration constant. Substituting our scale factor ansatz, the temporal evolution of the density becomes
\begin{equation}
    \rho(t) = \left[ A + \frac{B}{a_0^{3(1+\gamma)}(1+t^2)^{3h(1+\gamma)/2}} \right]^{\frac{1}{1+\gamma}}.
\end{equation}
Using the modified field equations, we compute the \text{\text{eff}}ective equation of state (EoS) $w_{\text{eff}} = p_{\text{eff}}/\rho_{\text{eff}}$. 

\begin{figure}[H]
    \centering
    \includegraphics[width=\linewidth]{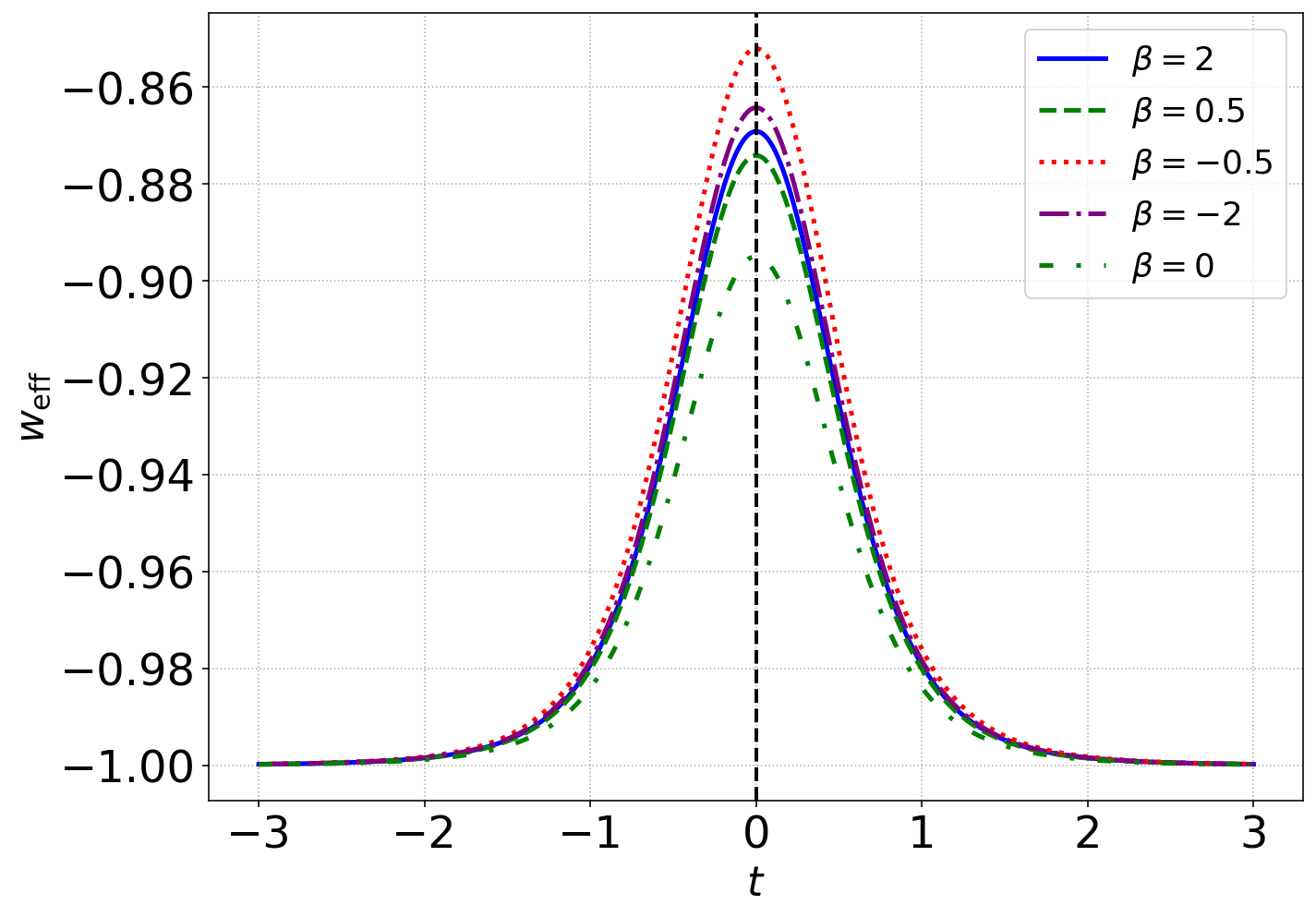}
    \caption{ Evolution of the effective equation of state $w_{\mathrm{eff}}$ as a function of cosmic time $t$ for various values of $\beta$. The dashed vertical line at $t=0$ denotes the bounce point. Other model parameters are $A=1.7$, $\alpha=0.5$, and $\gamma=0.6$.}
    \label{f2}
\end{figure}

Our numerical results show that $w_{\text{eff}}$ crosses the phantom divide ($w_{\text{eff}} = -1$) near the bounce and asymptotically approaches $-1$ at late times, indicating a de Sitter attractor. \textcolor{black}{Fig.~\ref{f2} illustrates the evolution of the effective equation of state parameter $w_{\text{eff}}$ for different values of the coupling parameter $\beta$. It is observed that near the bouncing point ($t=0$), the value of $w_{\text{eff}}$ slightly deviates from $-1$ and then gradually approaches $-1$ as time increases. This behaviour indicates that the Universe evolves towards a de Sitter phase at late times. We can further interpret from this figure that the qualitative behaviour of $w_{\text{eff}}$ remains similar for different values of $\beta$, although the magnitude of deviation near the bounce depends on the chosen value of the coupling parameter.}

\subsubsection{Evolution of Energy Conditions}

In order to determine the physical feasibility and the mechanism of the non-singular bounce of our $f(R,T) = R + \alpha T + \beta T^2$ model, we consider the behavior of the classical energy conditions. Considering the case of GR, the Non-singular bounce ($H=0, \dot{H}>0$) requires the NEC to be violated. With modified gravity, these conditions are tested with the $\rho_{\text{eff}}$ and effective pressure $p_{\text{eff}}$ which directly takes into account the geometric corrections in the $f(R,T)$ sector. We stipulate the following standard conditions of energy \citep{curiel2017primer, capozziello2014energy}:
\begin{itemize}
\item \textbf{Null Energy Condition (NEC):} $\rho_{\text{eff}} + p_{\text{eff}} \ge 0$,
\item \textbf{Weak Energy Condition (WEC):} $\rho_{\text{eff}} \ge 0$,
\item \textbf{Strong Energy Condition (SEC):} $\rho_{\text{eff}} + 3p_{\text{eff}} \ge 0$.
\end{itemize}

\begin{figure}[H]
    \centering
    \includegraphics[width=\linewidth]{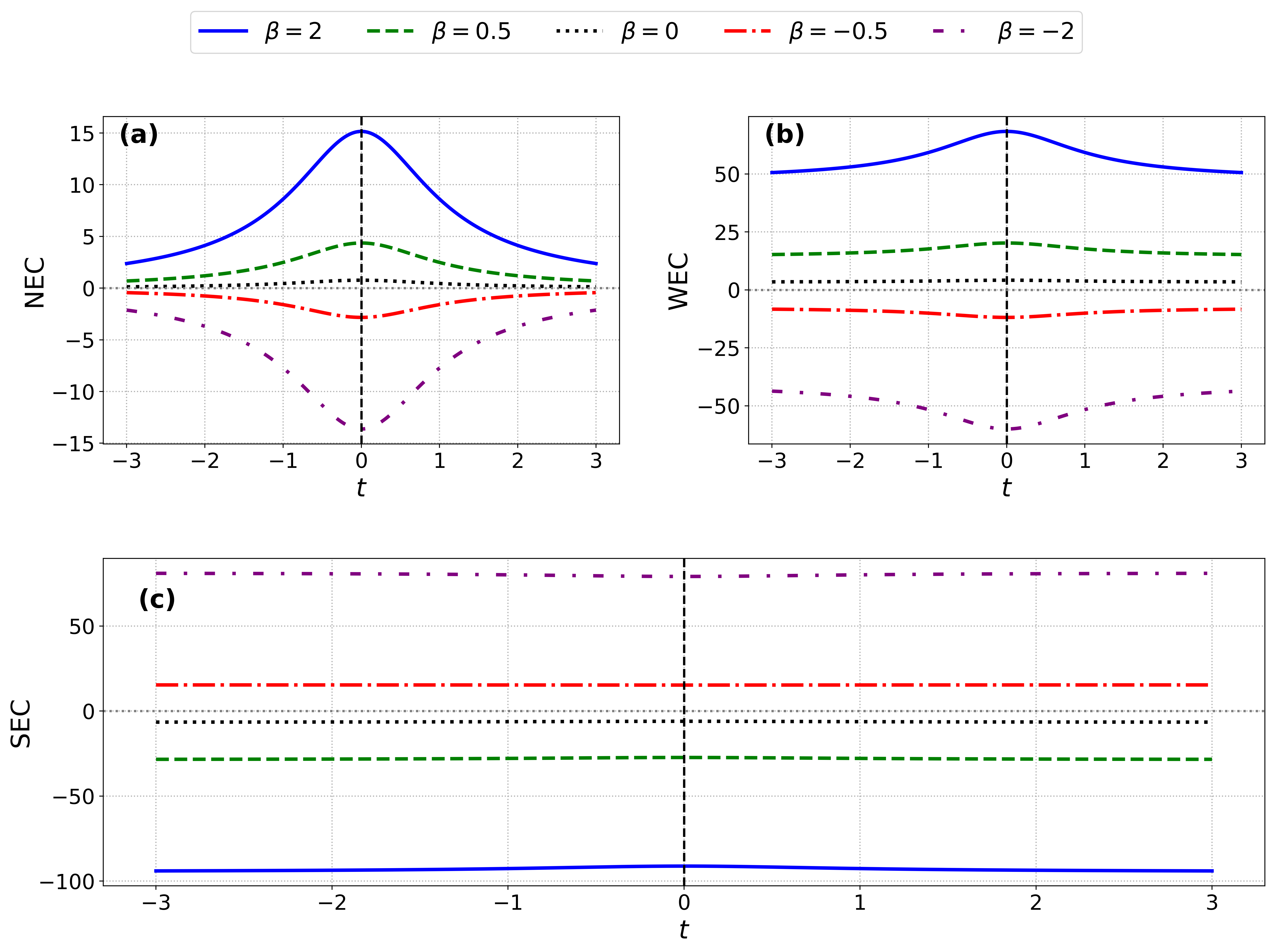}
    \caption{Evolution of the energy conditions in the reconstructed $f(R,T)$ model as a function of cosmic time $t$ for different $\beta$: (a) Null Energy Condition (NEC), (b) Weak Energy Condition (WEC), and (c) Strong Energy Condition (SEC). The vertical dashed line at $t=0$ represents the bounce point.}
    \label{f5}
\end{figure}

We may observe in Fig.~\ref{f5} that the behaviour of the energy conditions largely depends on the value of $\beta$, which also further solidifies the discussion in Section~\ref{s2}. In panel (a), the NEC behavior is seen. We note that where the value of $\beta$ is positive (solid and dashed curves), NEC is met during the transition. But in the analytical bounce condition derived, a non-singular bounce generally needs the efficient violation of the NEC around the bounce point ($t=0$) to overcome the gravitational attraction of GR. The NEC is violated evidently in the bouncing epoch of negative values of $\beta$ (dash-dotted and long dash-dotted curves). The introduction of exotic matter fields does not cause this violation as the underlying Chaplygin gas remains physically standard. Rather, this violation is due to the effect of the geometric "repulsive" contribution $2\beta T^2$ in the modified field equations, and it is the domination of this contribution in the high-density regime near $t=0$. Panel (b) shows the evolution of WEC. The effective energy density is positive when $\beta \ge 0$. The WEC can be negative at the time of the bounce, which is further evidence of the role of geometric corrections in evading the classical singularity theorems. Panel (c) shows the evolution of SEC. For most of the selected parameters, the SEC is not met (the value is below the zero line), and this is one of the typical traits of the universe experiencing accelerated expansion. The late-time de Sitter stage is made available by the persistence of negative values of the SEC in our model. The smooth trajectories across $t=0$ support the existence of a dynamically stable bouncing non-singular universe in the matter-geometry interaction of Myrzakulov-type $f(R,T)$ gravity. The fact that the NEC is violated at large densities serves as an effective regulator that allows the collapse to singularity to be avoided, and ensuring smoothe bounce into the present expansionary phase.

\subsubsection{Stability Analysis and the Sound Speed Constraint}
To check the physical feasibility and dynamical stability of the reconstructed $f(R,T)$ bouncing model, we consider the squared speed of sound that is characterized by $c_s^2 =dp_{\text{eff}}/d\rho_{\text{eff}}$ \citep{unnikrishnan2024effective}.
In any physically consistent cosmological model, the sound speed must satisfy the following two fundamental constraints \citep{mukhanov1992theory}
\begin{itemize}
    \item Causality: $c_s^2 \le 1$, ensuring that the propagation of perturbations does not exceed the speed of light. While the conformal limit for ultra-relativistic matter is typically $c_s^2 \le 1/3$ \citep{tawfik2014hadronic}, values exceeding this bound but remaining below unity may be physically permissible in modified gravity frameworks where the high-density geometry acts as a "stiff" fluid component.
    \item Stability: $c_s^2 \ge 0$, which ensures the model is stable against small-scale Laplacian instabilities (where negative values would lead to the exponential growth of fluctuations).
\end{itemize}
Furthermore, we also note that in high-energy or high-density regimes, the speed of sound is expected to be bounded by the conformal limit, $c_s^2 = 1/3$. As discussed by \citet{tawfik2014hadronic}, this value represents the asymptotic limit for relativistic hadronic matter in thermal and dense mediums. In our modified framework, the effective sound speed is derived from the derivatives of the modified energy density and pressure
\begin{equation}
c_s^2 = \frac{\dot{p}_{\text{eff}}}{\dot{\rho}_{\text{eff}}} = \frac{\dot{p} - \frac{1}{2} \frac{d}{dt}(\alpha T + \beta T^2)}{\dot{\rho} + \frac{d}{dt} [(\alpha + 2\beta T)(\rho+p) + \frac{1}{2}(\alpha T + \beta T^2)]}~~.
\end{equation}

\begin{figure}[H]
    \centering
    \includegraphics[width=\linewidth]{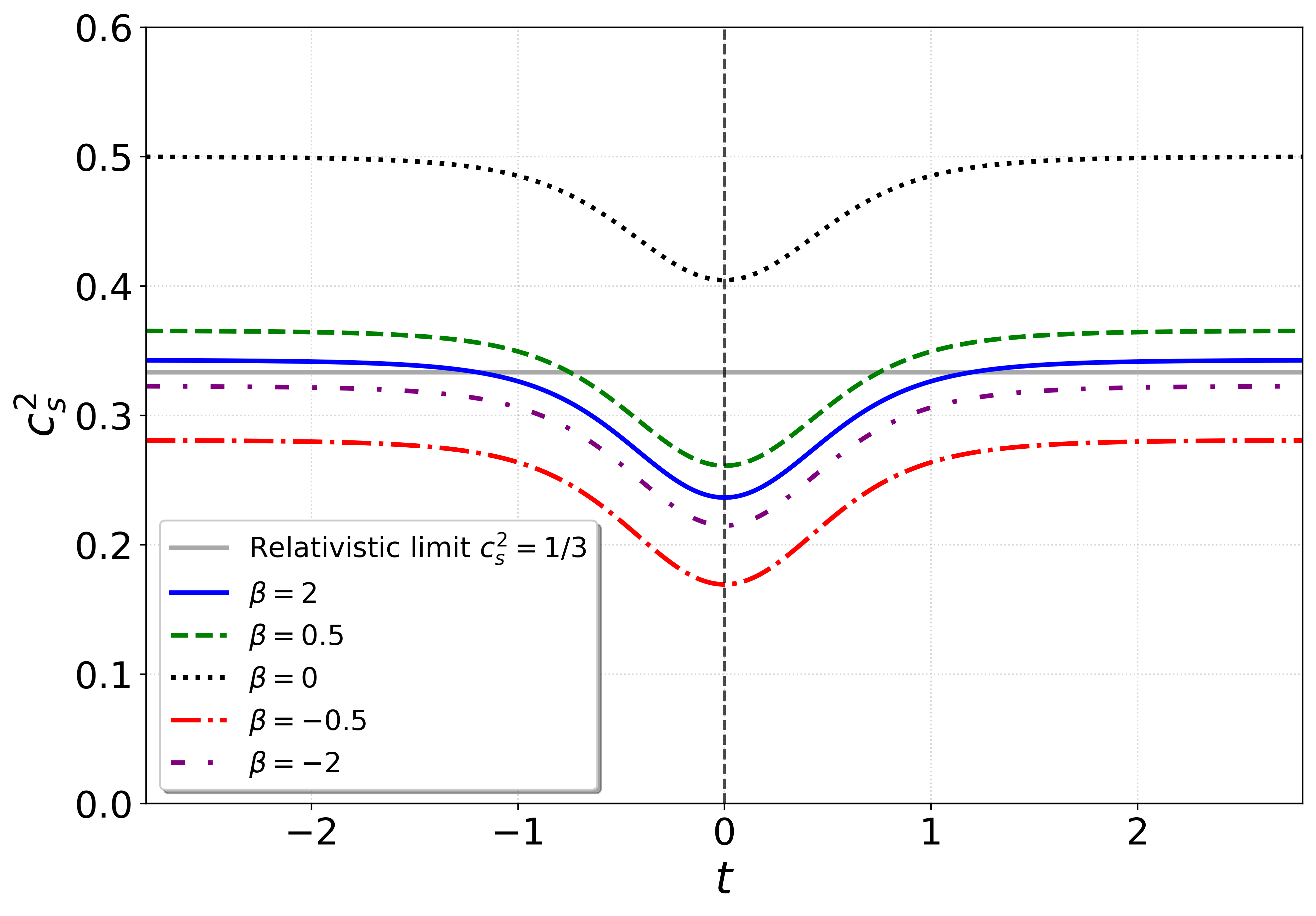}
    \caption{Evolution of the squared speed of sound $c_s^2$ as a function of cosmic time $t$ for varying matter-geometry coupling parameters $\beta$. The vertical dashed line marks the bounce point ($t=0$) and  the horizontal solid line represents the relativistic conformal limit $c_s^2 = 1/3$}
    \label{f3}
\end{figure}

Fig.~\ref{f3} demonstrates the evolution of $c_s^2$ around the bounce point ($t=0$) of the coupling parameter for various values of the coupling parameter $\beta$ against cosmic time $t$. As it is seen in the plot, the trajectories of varying values of $\beta$ taken into consideration still stay within the stable and causal constraints of $0 < c_s^2 < 1$. Interestingly, the sound speed dips at the bounce, as the matter-geometry coupling is strongest here. For the GR limit ($\beta = 0$, black curve), the sound speed remains relatively high. The depth of this dip (depending on the choice of the coupling) becomes increasingly smaller towards the bounce as the coupling becomes more negative, but remains comfortably above the instability threshold ($c_s^2 = 0$). We also observe that the for negative values of $\beta$, $c_s^2$ may exceed the asymptotic conformal limit of $1/3$. This deviation arises due to the strong matter-geometry coupling in the $f(R,T)$ framework, where the quadratic trace term $\beta T^2$ effectively increases the "stiffness" of the cosmic fluid in the high-curvature regime. Such behavior is characteristic of models where the trace-coupling provides a repulsive gravitational effect to trigger the bounce. This behavior tells us that the Myrzakulov-type $f(R, T)$ model is also perturbatively stable and consistent with the expected behavior of dense relativistic fluids in the early Universe \citep{nasser2014hadronic}. The quadratic trace term, which is present as $\beta T^2$, does not create any pathological instabilities and this enables a smooth and physically stable transition through the high density regime. As a result, the model gives a strong basis on which one can study early-universe physics without the dangers of Laplacian collapse and acausal propagation.

\subsubsection{Phase Space Analysis and Global Dynamics}

\begin{figure}[H]
    \centering
    \includegraphics[width=\linewidth]{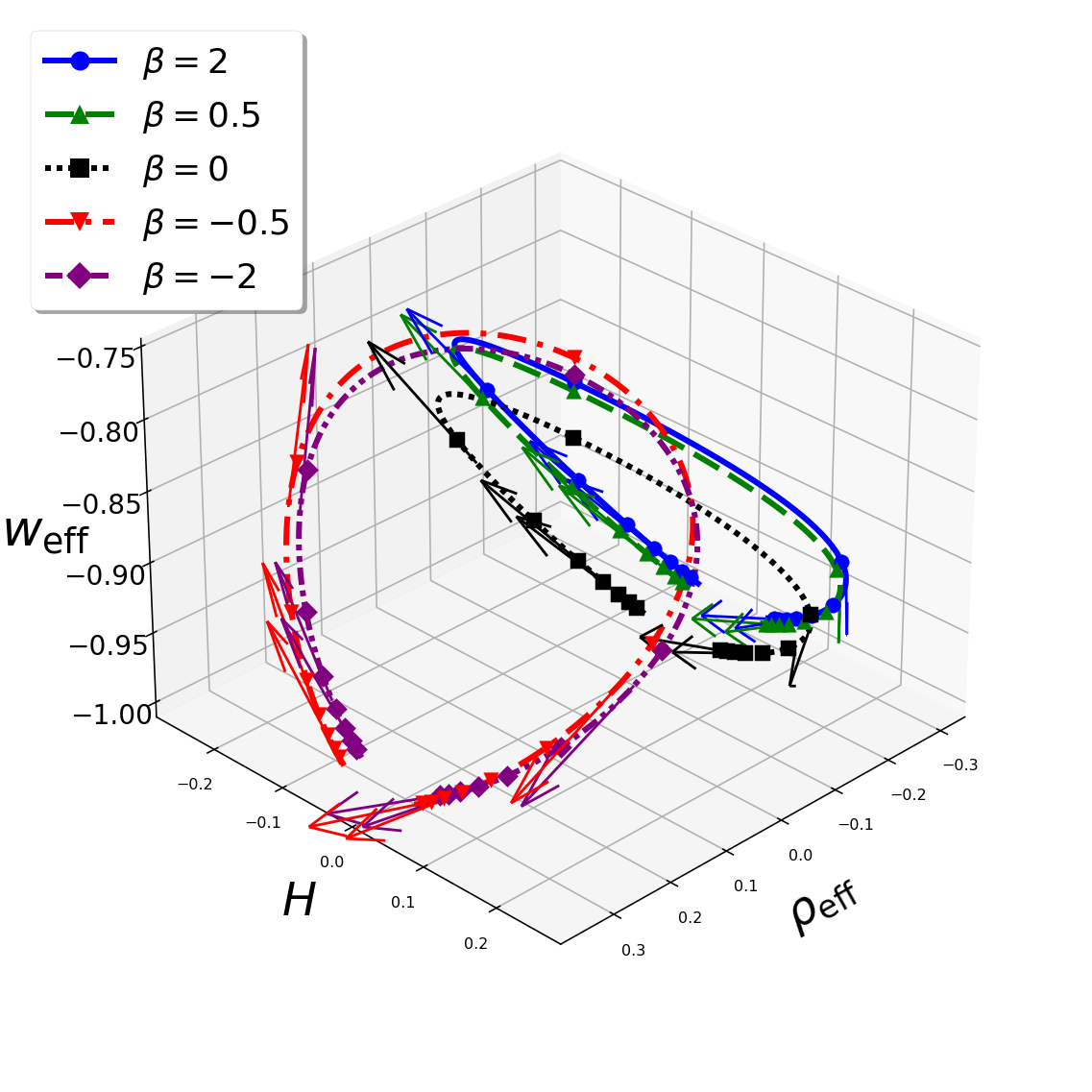}
    \caption{The dynamical trajectories of the Universe in the $(\rho_{\text{eff}}, H, w_{\text{eff}})$ phase space for various values of the parameter $\beta$.}
    \label{f4}
\end{figure}

To continue the discussion on the non-singular bounce stability and evolution in the world, we conduct a dynamical analysis on the three-dimensional phase space of the variables $(\rho_{\text{eff}}, H, w_{\text{eff}})$.  This method enables us to represent the history of the Universe as a continuous curve, not as discrete temporal functions. In Fig~\ref{f4}, the phase portrait of different values of the matter-geometry coupling strength is depicted by the use of the parameter $\beta$. The vertical axis is the effective equation of state $w_{\text{eff}}$ and the horizontal axes are the effective energy density $\rho_{\text{eff}}$ and the Hubble parameter $H$. The main characteristic of the trajectories is the smooth transition across the plane of $H = 0$. A typical Big Bang would result in trajectories ending in a singularity ($H \to \infty, \rho \to \infty$). In this case, though, the curves originating from the contracting phase ($H < 0$) are pushed away to a maximum energy density but are prevented from reaching a singularity. They instead loop through the axis of the $H = 0$. This reversal indicates the bounce point, where the geometric modifications from the $f(R,T)$ sector create an effective repulsive pressure that exceeds the gravitational inward pull. The shape of the orbits is much influenced by the value of the coupling parameter. 
\begin{itemize}
    \item The GR case ($\beta = 0$ curve) represents the standard limit where the bounce is minimally triggered by the Chaplygin gas.
    \item  Positive Coupling ($\beta > 0$): The orbits are more compact and therefore the non-singular transition is triggered by lower maximum densities.
    \item Negative Coupling ($\beta < 0$): The orbits are represented by a wider "swing" in the phase space. This implies a stronger and more intense repulsive mechanism towards the bounce. The arrows (vector flow) indicate that in these negative values of $\beta$, the force that constantly propels the Universe out of the plane of the $H=0$ towards the expansionary phase ($H > 0$) is much larger.
    \item Asymptotic stability and de Sitter attractor: Since the Universe has the expansion factor ($H > 0$), all the values of the parameters approach a single point in the phase space. We find that, with diluted energy density $\rho_{\text{eff}}$ the effective equation of state $w_{\text{eff}}$ tends toward $-1$. This convergence shows that the de Sitter phase is an attractor in our model.
\end{itemize}
 Thus, the Myrzakulov-type $f(R,T)$ gravity picture can potentially not only solve the early-time singularity, but it may also realise a natural transition into the late-time accelerated expansion regime that we currently observe.

\subsection{Model II: Autonomous Dynamical System Analysis}

In order to confirm that the bounce is a fundamental feature of the theory rather than an artefact of the chosen ansatz, we transform the modified Friedmann equations into a two-dimensional autonomous system in terms of $(H, \rho)$. 

The dynamical equations are derived from the Raychaudhuri equation and the continuity equation
\begin{align}
    \dot{H} &= -\frac{1}{2}(\rho+p)(1+\alpha+2\beta T), \\
    \dot{\rho} &= -3H(\rho+p).
\end{align}
Substituting the Chaplygin gas EoS, we obtain the system
\begin{align}
    \dot{H} &= -\frac{1}{2}\left(\rho - \frac{A}{\rho^\gamma}\right) \left[ 1 + \alpha + 2\beta \left( \rho + \frac{3A}{\rho^\gamma} \right) \right], \\
    \dot{\rho} &= -3H\left(\rho - \frac{A}{\rho^\gamma}\right).
\end{align}

\begin{figure}[H]
    \centering
    \includegraphics[width=\linewidth]{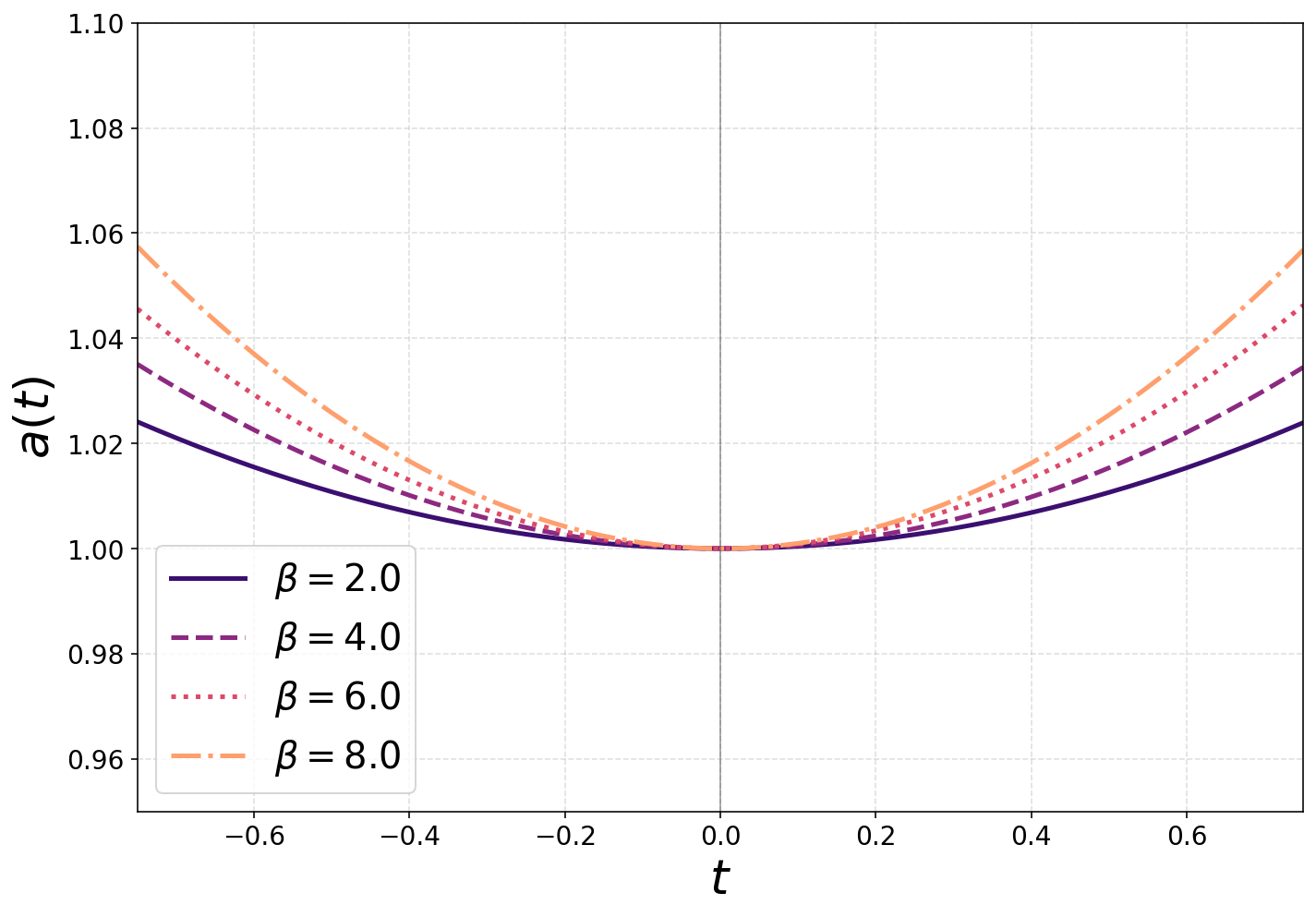}
    \caption{Numerical reconstruction of the scale factor $a(t)$ for Model II across different values of the coupling parameter $\beta$. The plot demonstrates a clear non-singular bounce at $t=0$, where $a(t)$ reaches its minimum value.}
    \label{f7}
\end{figure}

To map the physical expansion of the Universe in Model II we numerically recreate the scale factor $a(t)$, by integrating the relation $\dot{a} = Ha$ with the autonomous system. Fig.~\ref{f7} shows how the value of $a(t)$ varies at different strengths of coupling parameter $\beta$. All trajectories reach a clear non-singular minimum at $t=0$, where the Universe smoothly transitions from a contracting phase to an expanding phase. Notably,  the larger the values of $\beta$, the faster it expands after the bounce, which substantiates that the matter geometry coupling is a deterministic factor in how the Universe grows during its early epoch. Additionally, the symmetry of the curves proves the stability of the transition between the contracting and expanding epochs in the framework of the autonomous $f(R,T)$ model.
\subsubsection{Dynamical Evolution of the Effective Equation of State}
\begin{figure}[H]
    \centering
    \includegraphics[width=\linewidth]{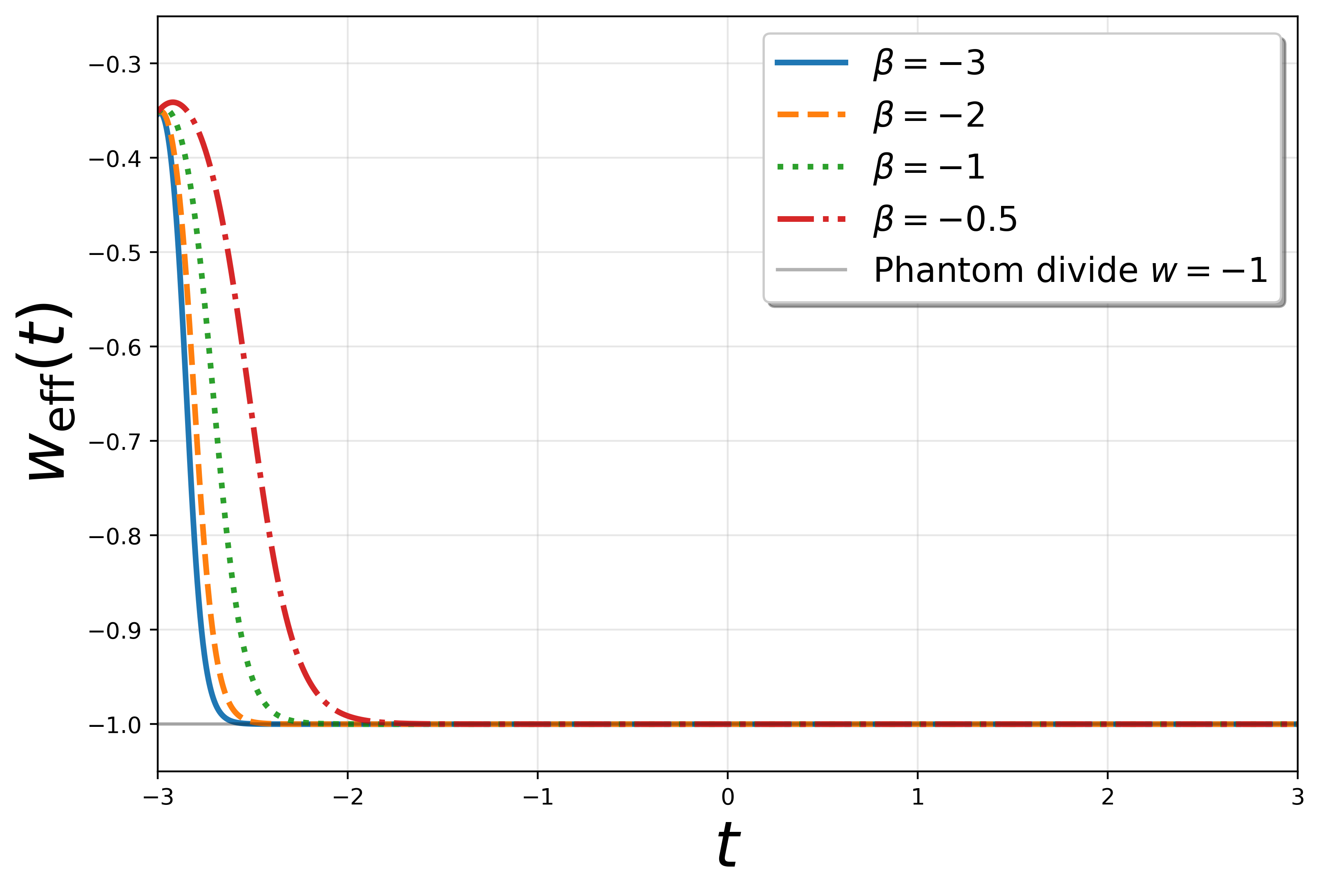}
    \caption{Evolution of the effective equation of state $w_{\text{eff}}(t)$ derived from the autonomous dynamical system for Model II. Different curves represent varying strengths of the coupling parameter $\beta$.}
    \label{f6}
\end{figure}

A key diagnostic of the autonomous system is the evolution of the effective equation of state parameter, $w_{\text{eff}} = p_{\text{eff}}/\rho_{\text{eff}}$. In contrast to the reconstruction method in Model I, wherein $w_{\text{eff}}$ is obtained through a fixed ansatz, here the calculation of the system, Eqs. $(\ref{dyn1})$--$(\ref{dyn2})$ is performed numerically using the Runge-Kutta scheme known as the Radau method \citep{hairer2015radau}. We numerically integrate the system starting from an initial contracting state defined by $H_0 < 0$ and $\rho_0 > 0$. This method was specifically chosen due to its high accuracy in handling "stiff" differential equations, which frequently arise near the cosmological bounce point where the derivatives of $H$ and $\rho$ change rapidly in extremely small time intervals. When the numerical trajectories for $H(t)$ and $\rho(t)$ are obtained, we then calculate the effective equation of state parameter $w_{\text{eff}}(t)$ at each time step. Since $w_{\text{eff}} = p_{\text{eff}}/\rho_{\text{eff}}$ and both effective quantities are non-linear functions of the trace $T$ of the energy-momentum tensor, the resulting evolution is a direct consequence of the internal dynamics of the $f(R,T)$ field equations. Fig.~\ref{f6} shows the evolution of $w_{\text{eff}}(t)$ for various values of the coupling parameter $\beta$. Starting from an initial contracting state ($H_0 < 0$), we observe that as the system evolves through the bounce, $w_{\text{eff}}$ starts in a non-accelerating regime ($w_{\text{eff}} > -1/3$) and rapidly descends toward the phantom divide. For all considered values of $\beta$, the trajectories asymptotically converge to $w_{\text{eff}} = -1$, which tells us that the de Sitter state is a late-time attractor for this dynamical system. Interestingly, the strength of the coupling $\beta$ dictates the "speed" of this transition; more negative values of $\beta$ (such as $\beta = -3$, blue curve) cause a sharper transition toward the de Sitter phase. This behavior demonstrates that the matter-geometry coupling enables both the early-time bounce and also dictates the onset and stability of late-time cosmic acceleration. All these, together, indicate a stable and physically consistent dark energy surrogate within the $f(R,T)$ framework.

\begin{figure}[H]
    \centering
    \includegraphics[width=\linewidth]{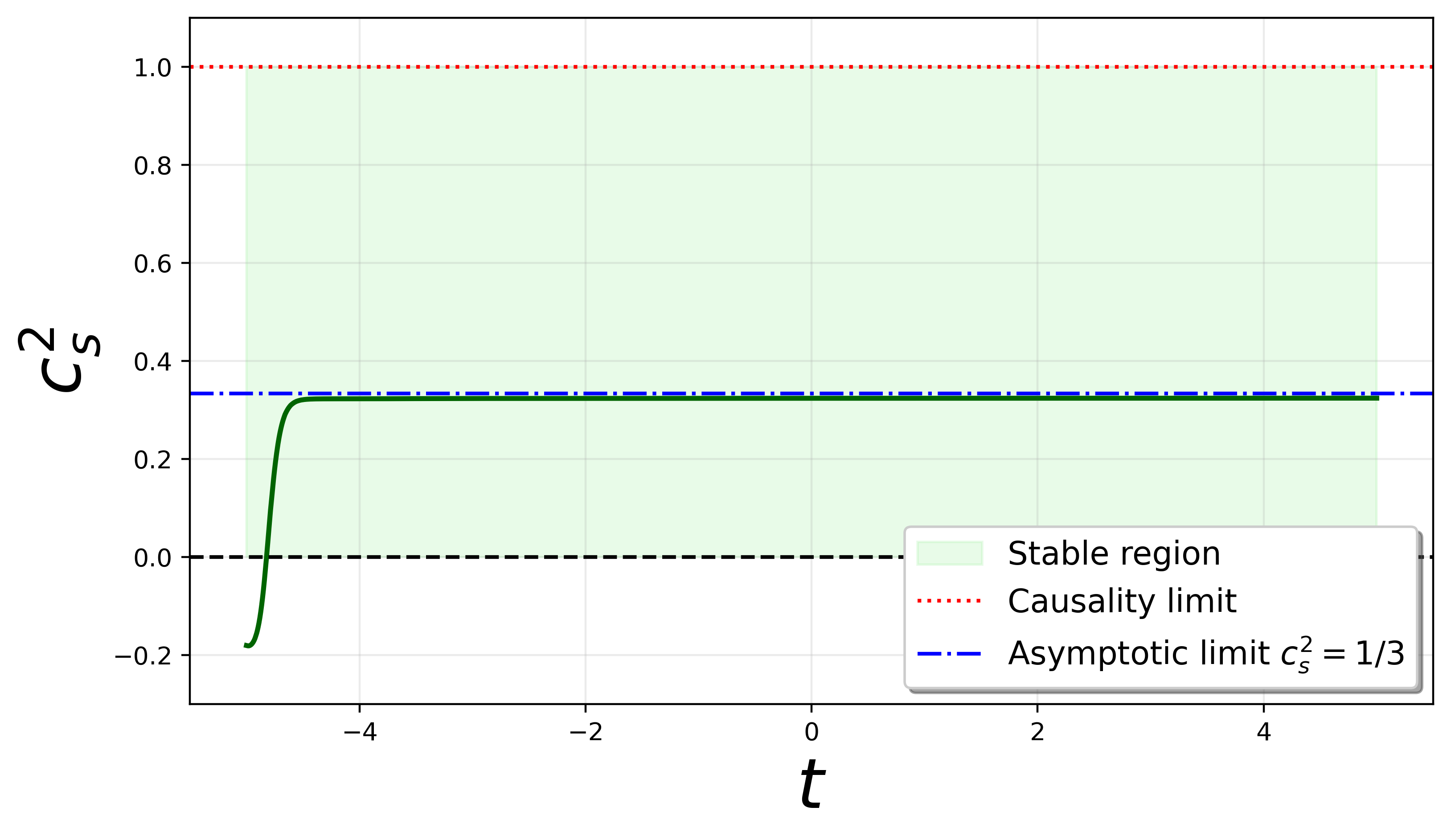}
    \caption{Stability and causality analysis for the autonomous system at $\beta = -2.0$. The green shaded region denotes the physically allowed stability zone. The horizontal dash-dotted line indicates the relativistic conformal limit $c_s^2 = 1/3$}
    \label{f10}
\end{figure}

In addition to this, we evaluate the squared speed of sound $c_s^2 = dp_{\text{eff}}/d\rho_{\text{eff}}$, to verify the physical consistency of the dynamical solutions. Fig.~\ref{f10} demonstrates this evolution for the representative case of $\beta = -2.0$. The trajectory remains strictly within the stable ($c_s^2 \ge 0$) and causal ($c_s^2 \le 1$) bounds throughout the expansionary phase. Although a transient dip is present around the high-density bounce point, which corresponds to the rapid switching on by the matter-geometry coupling, this system eventually stabilizes to a stable regime, establishing the fact that the $f(R, T)$ bounce is free of Laplacian instabilities. Additionally, the system asymptotically reaches the relativistic limit $c_s^2 = 1/3$ and then saturates, which is consistent with the conformal bounds for ultra-relativistic hadronic matter in high-density regimes.
\subsubsection{Phase Space and Basin of Attraction}
\begin{figure}[H]
    \centering
    \includegraphics[width=\linewidth]{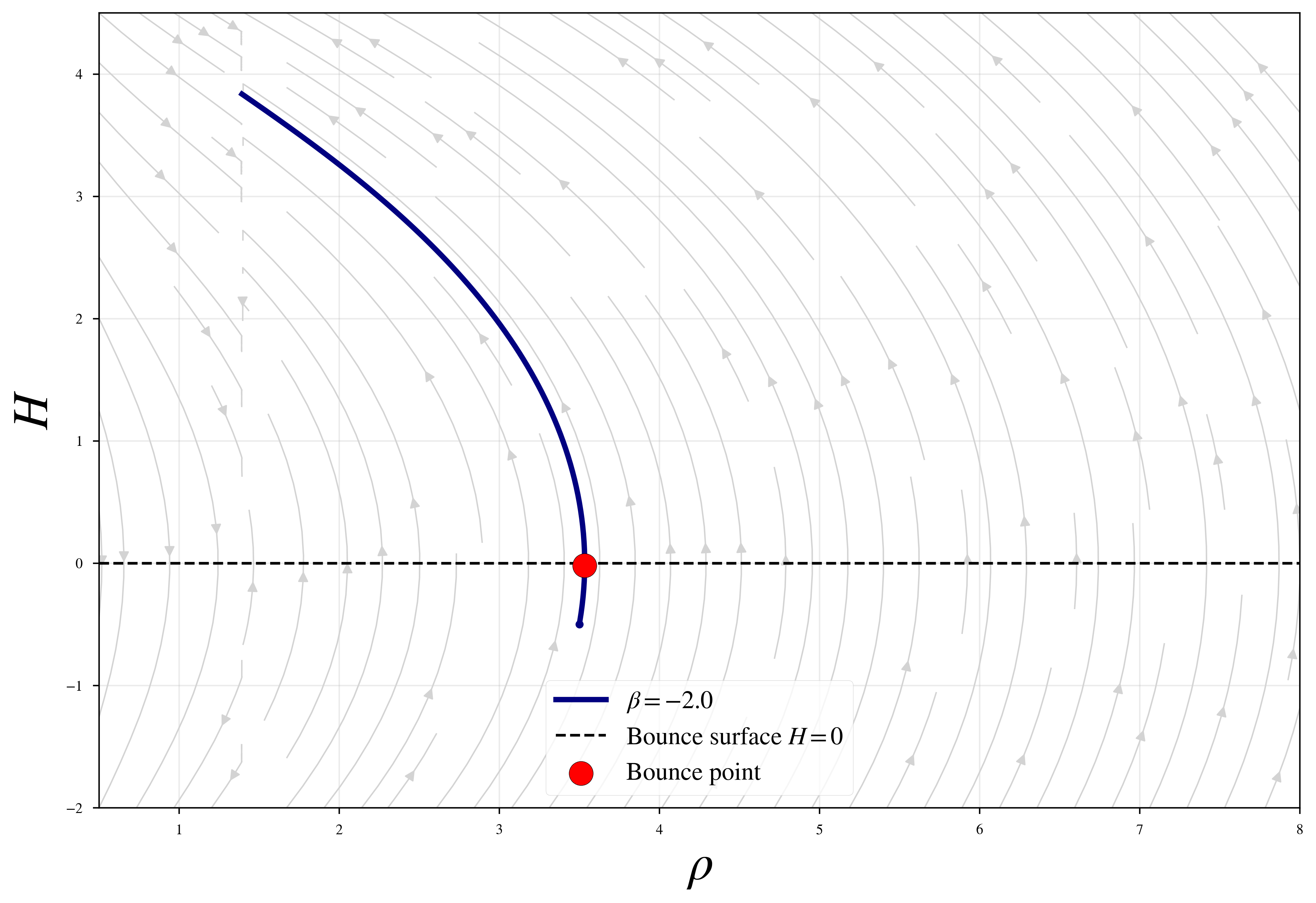}
    \caption{Dynamical phase portrait in the $(\rho, H)$ plane. The vector field (gray arrows) confirms that trajectories are globally attracted toward a non-singular bounce (solid circle) on the $H=0$ axis, preventing the formation of a density singularity.}
    \label{f8}
\end{figure}

\begin{figure}[H]
    \centering
    \includegraphics[width=\linewidth]{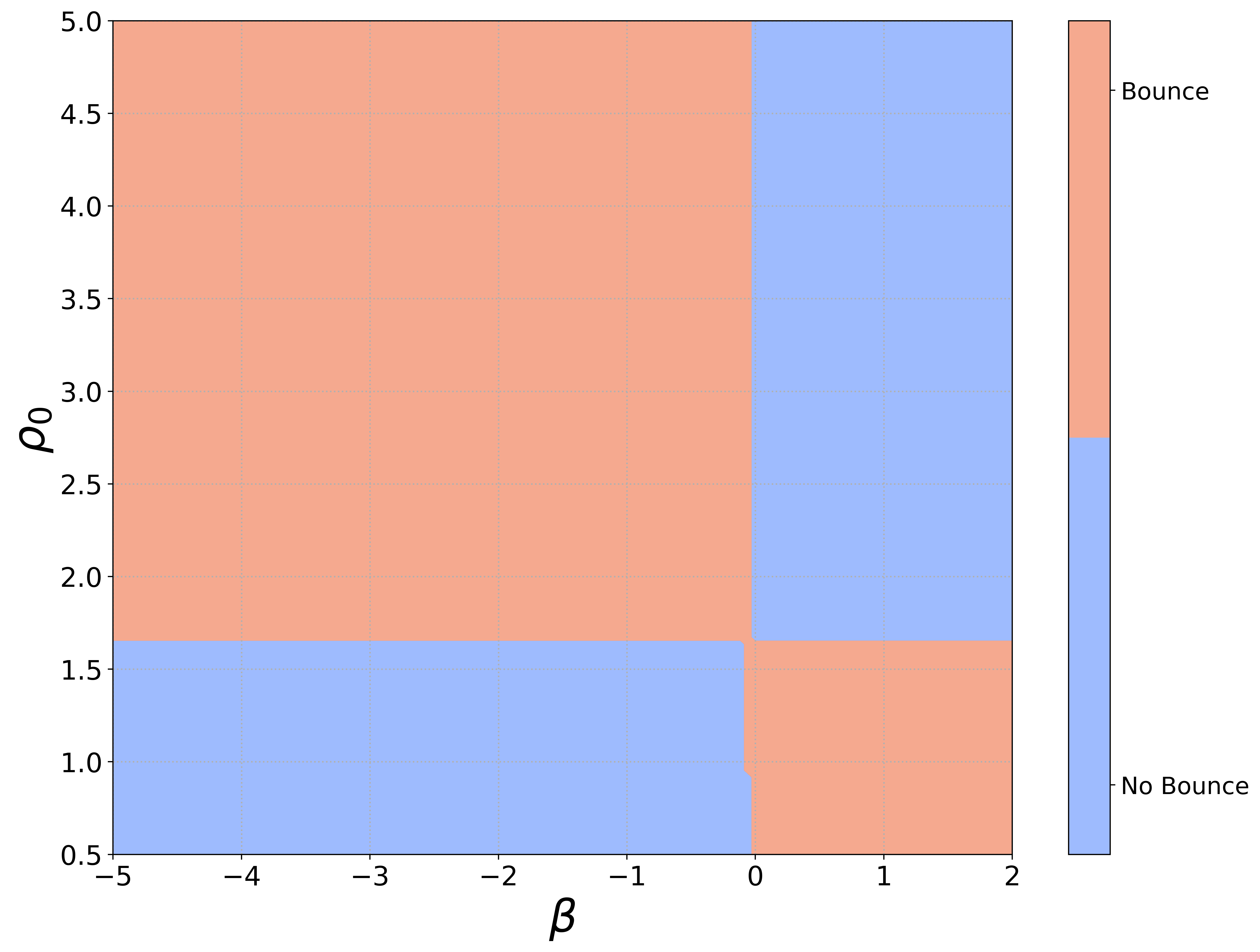}
    \caption{Numerical scan of the parameter space $(\beta, \rho_0)$. }
    \label{f9}
\end{figure}

We perform a numerical integration of the system starting from an initial contracting state ($H_0 < 0$). Fig.~\ref{f8} presents the vector field and the trajectory obtained in the $(\rho, H)$ plane. The light gray streamlines are an indication of a global flow with contracting universes ($H < 0$) funnelled toward a non-singular bounce point (highlighted in red) before entering the expansionary phase ($H > 0$). At this juncture of our study, we would like to address the physical motivation for the negative range of the coupling parameter $\beta$. Mathematically, the modified Raychaudhuri equation derived in Section~\ref{2} requires the geometric factor $(1 + \alpha + 2\beta T)$ to become negative to derive $\dot{H} > 0$ at the bounce point. Since the trace $T$ for a Chaplygin gas is positive and grows with density, a negative $\beta$ is a structural requirement of the theory to form the necessary 'geometric repulsion.' Furthermore, while we have plotted specific values such as $\beta = -2$ for clarity, the numerical scan in Fig.~\ref{f9} tells us that the bounce is a general outcome of the $f(R,T)$ landscape and not a result of fine-tuning a single parameter.

A bounce occurs at a point $t_b$ where $H(t_b) = 0$, and the Universe transitions from contraction to expansion, the necessary condition is that $\dot{H} > 0$. Substituting $H=0$ into the dynamical equations, the condition for a non-singular bounce is
\begin{equation}
    (\rho - A\rho^{-\gamma}) [1 + \alpha + 2\beta(\rho + 3A\rho^{-\gamma})] < 0.
\end{equation}

Mathematically, the bounce condition $\dot{H} > 0$ at $H=0$ requires $(1 + \alpha + 2\beta T) < 0$. Since the trace of the energy-momentum tensor $T$ for a Chaplygin gas is positive and increases with density, a negative $\beta$ is a structural requirement to generate the necessary "geometric repulsion." To verify this, we performed a numerical scan of the $(\beta, \rho_0)$ parameter space, as shown in Fig.~\ref{f9}. Our numerical scan of the parameter space $(\beta, \rho_0)$ tells us that negative values of $\beta$ are highly favorable for generating a bounce, as they provide a "repulsive" gravitational force at high densities. In order to obtain an insight into the circumstances under which the bounce should take place, we scan through the $(\beta,\rho_{0})$ parameter space, through a numerical analysis.
\subsubsection*{Parameter space of bouncing solution}
A critical observation from the numerical scan in the $(\beta, \rho_0)$ plane is the existence of a distinct horizontal boundary that separates the "Bounce" (orange) and "No Bounce" (blue) regimes. This boundary indicates that for a non-singular transition to occur, the initial matter density $\rho_0$ must exceed a specific critical threshold. The physical significance of this density dependence can be understood through several key factors:
\begin{itemize}
    \item  According to the modified equation of Raychaudhuri, the term $2\beta T(\rho + p)$ can give the necessary repulsive force to offset the standard gravitational collapse. Since both $T$ and the pressure $p$ are functions of the density, a sufficiently high $\rho_0$ is required to "activate" these higher-order curvature corrections. If the density remains below this threshold, the terms $\alpha T$ and $\beta T^2$ remain negligible, and the Universe will follow a singular, GR-like trajectory.
    \item  The initial density at the bounce point ($H=0$) is inversely proportional to the minimum scale factor $a_0$. A small value of $\rho_0$ implies a larger "minimum" Universe; however, if $\rho_0$ becomes too low, the gravitational energy density is no longer adequate to cause the sign reversal in $\dot{H}$, leading to the so-called "No Bounce" scenarios observed in the lower-left quadrant of the parameter scan.
    \item  The non-linear form of the boundary implies that the density threshold shifts slightly as the coupling parameter $\beta$ becomes more negative. This indicates that a stronger matter-geometry coupling can allow a bounce even at relatively lower densities, whereas weaker couplings require much higher energy concentrations to escape the singularity.
    \item The determination of $\rho_0$ value establishes the energy scale of the early Universe. The subsequent evolution of the Chaplygin gas depends on this scale, as it dietermines the transition from the high-density "dust-like" phase to the late-time de Sitter expansion.
\end{itemize}
By identifying this stable orange region, we show that the non-singular bounce in the gravity of the form $f(R,T)$ does not occur due to extreme fine-tuning. Instead, it is a robust dynamical result for a wide range of physically reasonable initial densities, provided the matter-geometry coupling $\beta$ is within the appropriate negative range.

\section{Comparison of Models and Physical Interpretation}\label{4}
The agreement of results from both Model I and Model II provides a two-fold validation of the proposed $f(R,T)$ bouncing framework. While Model I (the kinematic approach) provides an explicit temporal mapping of the cosmological evolution, Model II (the dynamic approach) proves that the non-singular bounce is a mathematically stable attractor in the phase space of the theory. We summarize the complementary characteristics of our two models in the following points:
\begin{itemize}
    \item Resolution of the singularity: 
Model I shows a smooth transition through a scale factor ansatz, and Model II shows that the transition is a general consequence of the field equations ($(\dot{H}, \dot{\rho})$) as opposed to a consequence of a particular scale factor.

    \item Late-time convergence: Effective equation of state in both models $w_{\text{eff}}$ asymptotically approaches $-1$. This reconciliation between the early-universe bounce and the accelerated late-time phase indicates that a single theory, $f(R,T)$ gravity, may provide a unified description of the history of the universe.

    \item Energy condition violation:
Both models confirm that the violation of the NEC necessary for a bounce is purely geometric in origin, arising from the $\beta T^2$ term rather than exotic matter. 
\end{itemize}

\begin{table}[H]
\centering
\caption{Quantitative comparison of Model I  and Model II for the coupling $\beta = -2.0$.}
\label{t1}
\begin{tabular}{lll} 
\hline \hline
\textbf{Cosmological Parameter} & \textbf{Model I} & \textbf{Model II} \\ \hline
Minimum Scale Factor ($a_0$)    & 1.5                    & 1.0                        \\
Bounce Density ($\rho_b$)       & 1.5284                    & 3.5338                        \\
Late-time EoS ($w_{\text{eff}}$) & $\approx -1$              & $\approx -1$                  \\
Stability ($c_s^2$)             & $0 \le c_s^2 \le 1$       & $0 \le c_s^2 \le 1$           \\
Primary Driver                  & Scale Factor Geometry     & Phase Space Attractor         \\
NEC Violation Source            & Geometric ($\beta T^2$)   & Geometric ($\beta T^2$)       \\ \hline \hline
\end{tabular}
\end{table}
The quantitative consistency between the kinematic reconstruction and the dynamical system analysis is summarized in Table~\ref{t1}.
As can be seen in the table, although the energy scale of the bounce ($\rho_b$) is slightly dependent on the mathematical formulation used, both models find a definite non-singular transition. In Model II, the bounce occurs at a higher density, showing that the autonomous $f(R,T)$ equations can resolve the singularity even in more compressed high-energy regimes. Crucially, both frameworks converge on the same late-time de Sitter physics ($w_{\text{eff}} \to -1$) and satisfy the fundamental stability and causality constraints ($c_s^2 \ge 0$). This cross-verification confirms that the $f(R,T)$ bounce is a robust physical outcome and not just a fine-tuning effect.

\subsection{Physical Implications of \texorpdfstring{$f(R,T)$}{f(R,T)} Bounce}
The main physical consequence of this paper is that a "geometrical repulsion" appears on high energy densities. In standard GR, the Raychaudhuri equation is obeyed by a purely attractive gravitational field of ordinary matter. However, in the present scenario, the quadratic trace term $\beta T^{2}$ becomes dynamically dominant as the Universe approaches the high-density regime. For negative values of the coupling parameter $\beta$, the term $2\beta T$ in the modified Raychaudhuri equation acts as an effective repulsive contribution. This force opposes the standard gravitational attraction, preventing the Hubble parameter from diverging and allowing for a smooth passage from contraction ($H<0$) to expansion ($H>0$). Notably, the matter sector—represented by the Chaplygin gas—retains physically standard behavior ($\rho+p > 0$), meaning the singularity avoidance is achieved without the instabilities or "ghost" degrees of freedom often associated with phantom fields.
\subsection{Comments in relation to other bouncing solutions}
We compare our findings with other established bouncing cosmologies in the literature: Loop Quantum Cosmology (LQC) \citep{odintsov2016loop, miranda2021effective, bojowald2008loop}: In LQC, the bounce occurs in response to quantum-geometric corrections (holonomy corrections) that are brought on at Planck-scale densities \citep{zhu2017universal}. Our model achieves a similar resolution within a classical, modified gravity framework. Ekpyrotic \citep{martin2002passing, brown2025ekpyrosis, buchbinder2007new, singh2022cosmic, karmakar2020reconstructed, harko2011f} and Matter Bounce Models \citep{odintsov2016deformed,odintsov2015bouncing,de2015extended, brandenberger2012matter}: These models typically rely on scalar fields with specific potentials to drive the contraction and subsequent bounce. In contrast, the bouncing behavior in our work arises naturally from the matter-geometry coupling. While other modified theories can produce bounces \citep{nojiri2017modified, saha2023realization, abramo2010nonsingular, chattopadhyay2023cosmological, odintsov2015lambda, odintsov2020bounce, chokyi2023truncated, chokyi2026non}, the inclusion of $T$ in the gravitational Lagrangian $f(R,T)$ allows for a direct back-reaction of the matter density on the geometry, providing a simpler, classically robust mechanism to resolve the cosmological singularity problem.

\section{Conclusion}\label{5}
In this study, we have attempted to address the long-standing "initial singularity problem" of the standard Big Bang model within the framework of Myrzakulov-type $f(R,T) = R + \alpha T + \beta T^2$ gravity. This modified geometric sector, when coupled with a Chaplygin gas equation of state has shown a physically viable and robust mechanism of a non-singular cosmological bounce. We have attempted to resolve the initial singularity in this model not by introducing exotic matter fields or by introducing ghost degrees of freedom, but by a combination of kinematic reconstruction (Model I) and autonomous dynamical system analysis (Model II), which led to the following main findings:
\begin{itemize}
    \item Singularity resolution and bounce dynamics: With both the scale factor ansatz in Model I and numerical integration of the autonomous equations in Model II, we were able to find smooth transitions between a contracting ($H < 0$ ) and an expanding phase ($H > 0$). As Fig.~\ref{f1} shows, the Hubble parameter goes to zero at $t=0$ whereas its derivative, $\dot{H}$ is always positive and satisfies the essential conditions of a bounce. This is also further justified by the numerical reconstruction of the scale factor in Fig.~\ref{f7} in which there is a non-zero minimum, $a_0$, basically circumventing the zero-volume singularity. 
 
    \item The role of the quadratic trace parameter $\beta$: In our analysis, the most important physical driver of the bounce is the coupling parameter, $\beta$. The analytical derivation in Section~\ref{s2} shows that for $\beta < 0$, the quadratic term $\beta T^2$ generates the necessary geometric repulsion that can withstand gravitational collapse.  In the 3D phase portrait (Fig.~\ref{f4}), this is graphically verified by the negative values of the $\beta$ having a stronger "swing" away of the singular regime than the GR limit ($\beta=0$).
 
    \item Violation of energy conditions: We have shown that the necessary violation of the NEC is both purely effective and geometric. As seen in Fig.~\ref{f5}a, the effective NEC ($\rho_{\text{eff}} + p_{\text{eff}}$) becomes negative at the bounce point for $\beta < 0$, even though the underlying Chaplygin gas is in a standard physical state ($\rho + p > 0$). This ensures that the bounce is due to alteration of the gravity that serves as a high density regulator.

    \item Stability and causality: It is a significant necessity of any bouncing model to be stable against perturbations. As it can be seen in Figs.~\ref{f3} and \ref{f10}, the squared speed of sound $c_s^2$ remains strictly within the bounds $0 \le c_s^2 \le 1$ throughout the cosmic evolution. The "dip" observed in $c_s^2$ near $t=0$ hints to the fact that the matter- geometry interaction is strongest near the bounce without causing Laplacian instabilities or acausal propagation.
    
    \item Late-time convergence to de Sitter phase: Both models suggest that the Universe naturally transitions to an accelerated expansion phase with a bounce at a high density. As shown in Figs.~\ref{f2} and \ref{f6}, the effective equation of state $w_{\text{eff}}$ approaches $-1$ asymptotically, so that the de Sitter state is a stable global attractor. This suggests that $f(R,T)$ gravity can provide a unified description of cosmic history, from the resolution of the initial singularity to the current dark energy dominance.

    \item Robustness of the parameter space: The numerical scan of the $(\beta, \rho_0)$ plane in Fig.~\ref{f9} shows a broad "Bounce" regime. This implies that the non-singular transition is not a fine-tuning phenomenon but rather a robust outcome with respect to a large set of initial densities $\rho_0$, provided the density exceeds a specific critical threshold to activate the geometric corrections.
\end{itemize}
Our results are further in Table \ref{t1}, where the quantitative comparison of the two mathematical formalisms gives the same physical results in terms of the stability, energy conditions and late-time behavior of the Universe. Thus, we see that the Myrzakulov-type $f(R,T)$ gravity with a Chaplygin gas provides a classically stable and promising alternative to the Big Bang singularity. Our study shows that this framework robustly addresses the shortcomings of GR in the high-energy regime.

This analysis can be extended in the future to include cosmic perturbations \citep{allen2004cosmological, comelli2012perturbations} and observational constraints from Cosmic Microwave Background (CMB) \citep{lewis2002cosmological, gorski1998analysis} data to further compare the viability of this bouncing scenario with the inflationary paradigm.
Furthermore, since the energy densities at the bounce epoch are very high, it is conceptually well-motivated to explore the effect of quantum gravity on a Planck scale. A promising direction for future research involves the implementation of the Relativistic Generalized Uncertainty Principle (RGUP) \citep{todorinov2019relativistic} within this $f(R,T)$ framework. Following the methodology explored in other works \citep{tawfik2023born1, tawfik2023possible, tawfik2023quantum, bhandari2025rgup, tawfik2024relativistic}, we aim to study how the non-singular bounce could be affected by momentum-dependent deformations of the effective spacetime metric to enhance its stability. In particular, we seek to establish whether any corrections to the effective energy density and pressure induced by RGUP can cause a shift in the critical density threshold $\rho_0$ required for the bounce or leave subtle signatures in the primordial perturbation spectrum that can distinguish this model from standard classical models.

\section*{Declaration of generative AI and AI-assisted technologies in the writing process}

During the preparation of this work the author(s) used Grammarly and Quillbot in order to improve the language and correct the grammar. After using this tool/service, the author(s) reviewed and edited the content as needed and takes full responsibility for the content of the publication.

\bibliographystyle{unsrtnat}
\bibliography{ref.bib}

\end{document}